\documentclass[aps,preprint,pra]{revtex4}

\usepackage{graphicx}

\usepackage{multirow}
\usepackage{color}

\begin{document}
\title{Topological phase in oxidized zigzag stanene nanoribbons}
\author{Mohsen Modarresi$^1$, Wei Bin Kuang$^2$, Thaneshwor P. Kaloni$^2$, Mahmood Rezaee Roknabadi$^1$, Georg Schreckenbach$^2$}
\email{Georg.Schreckenbach@umanitoba.ca}
\affiliation{$^1$Department of Physics, Ferdowsi University of Mashhad, Mashhad, Iran}
\affiliation{$^2$Department of Chemistry, University of Manitoba, Winnipeg, MB, R3T 2N2, Canada}

\begin{abstract}
First-principles and semi-empirical tight binding calculations were performed to understand the adsorption of oxygen on the surface of two dimensional (2D) and zigzag stanene nano-ribbons. The intrinsic spin-orbit interaction is considered in the Kane-Mele tight binding model. The adsorption of an oxygen atom or molecule on the 2D stanene opens an electronic energy band gap. We investigate the helical edge states and topological phase in the pure zigzag stanene nano-ribbons. The adsorption of oxygen atoms on the zigzag stanene nano-ribbons deforms the helical edge states at the Fermi level which causes topological (non-trivial) to trivial phase transition. The structural stability of the systems is checked by performing $\Gamma$-point phonon calculations. The adsorption of an oxygen atom or molecule on the 2D staneneSpecific arrangements of adsorbed oxygen atoms on the surface of zigzag stanene nano-ribbons conserve the topological phase which has potential applications in future nano-electronic devices.  
\end{abstract}
\maketitle
\section{Introduction}
After the discovery of graphene \cite{R1}, other 2D nano-structures composed of group IV honeycomb lattices were theoretically proposed and synthesized \cite{R2,R3,R4,R5}. Topological insulators were observed experimentally for 3D nano-structures \cite{R6,R7,R8,R9} and predicted in low buckled 2D nano-structures \cite{R11,R12,R13,R14,R15,sr,prb14,jpcc14,apl14,pssrrl1,pssrrl2}. The condition required for observing the Hall effect is to break the time-reversal invariance by applying a strong magnetic field. Kane and Mele modeled the intrinsic spin-orbit interaction (SOC) in 2D graphene and report a quantum Anomalous Hall effect for different spin directions which is called the quantum spin Hall effect or $Z_2$ topological insulator \cite{R14,R16}. Graphene is not an appropriate candidate for experimental realization of the quantum spin Hall effect because of a very weak intrinsic spin-orbit interaction ($10^{-3}$ meV)\cite{R17,R18}. By increasing the atomic mass using heavier atoms the intrinsic spin-orbit interaction strength is increased. 

Stanene is a cousin of graphene; it is a hexagonal buckled arrangement of Sn atoms in 2D. Stanene has been synthesized using molecular beam epitaxy \cite{R2}. The relatively strong spin-orbit interaction magnitude in stanene opens an electronic gap around 70 meV and 0.3 eV in pure and halogenated stanene, respectively \cite{R19}. Theoretical calculations based on density functional theory (DFT) predict topological phases in pure, halogenated \cite{R19} and hydrogenated \cite{R11} stanene. Due to the buckled structure a perpendicular external electric field produces a staggered sub-lattice potential that modifies the energy band gap \cite{R20} and removes the topological phase of 2D nano-structures \cite{R21,R22,R23,R24}. Also a critical value of applied strain causes trivial to topological phase transition\cite{R12,R25,R26}. 

In the topological phase, surface states between two time-reversal invariant momenta (TRIM) points intersect Fermi energy in an odd number of times \cite{R27} that guarantees the robustness of edge states against weak disorder. Based on previous study, the $Z2$ topological invariant can be extracted by analyzing the parity of the occupied wave function in the structure with inversion symmetries \cite{R28}. Also Soluyanov and Vanderbilt proposed a method based on the time-reversal polarization for computing topological invariant without inversion symmetry\cite{WIN}. The $Z2$ topological phase of 2D functionalized stanene was studied in previous works \cite{R13,R19}, but to the best of our knowledge the adsorption of different atoms on zigzag stanene nano-ribbons has not been reported. Due to the possibility of oxidation of fabricated nano-devices, one needs to address the adsorption of oxygen on two dimensional nano-structures for possible future applications. 

Graphene oxide \cite{R29,R30,jmc,apl1,apl2} and silicene oxide \cite{R31,R32} were investigated theoretically and experimentally before. Motivated by these works, here we present a study of oxygen adsorption on 2D stanene using DFT and tight binding approaches. The tight binding parameters are extracted by fitting two models. The obtained tight binding parameters are used to study the effect of oxygen adsorption on the topological phase in zigzag stanene nano-ribbons.  

\section{Model and method}
Our calculations include an ab-initio study of oxygen adsorption on the 2D structure of stanene and tight binding modeling of oxygen atom adsorption on zigzag nano-ribbons with different widths. In the first part we perform DFT calculations and fit these with tight binding results for the 2D structure to obtain required parameters. The on-site energy, spin-orbit interaction strength and hopping between nearest neighbor atomic sites for describing the tight binding Hamiltonian of a monolayer of pure stanene are extracted from our previous work \cite{R20}. The remaining parameters which include the on-site energy of oxygen atoms and hopping between tin and oxygen atoms are obtained in the present work.

\subsection{Density functional theory}
All the DFT calculations were performed by employing the Quantum-ESPRESSO package \cite{QE}; the generalized gradient approximation (GGA) and the Perdew-Burke-Ernzerhof (PBE) \cite{RPBE} exchange correlation functional were adopted. To avoid interaction between adjacent stanene layers in neighboring unit cells, a vacuum layer of 14 \AA  was used. An $8\times8\times1$ Monkhorst Pack k-point mesh was adopted to sample the 2D hexagonal Brillouin zone and the cut-off energy for the plane wave functions was set to 550 eV in all calculations. We used ultra soft and full relativistic pseudo-potentials to include SOC in the DFT calculations. Through geometry optimizations, cell parameters as well as atomic positions were fully relaxed until the forces on the atoms were less than 0.002 eV/\AA. We consider a $4\times4$ supercell of stanene that includes 32 Sn and various of O atoms. Different atomic positions for the adsorption of an oxygen atom and molecule were examined to find the minimum energy position for oxygen. Fig.\ 1 shows the optimized structure of pure stanene and adsorption of an oxygen molecule and atom on the 2D stanene. The adsorption of oxygen on the surface of stanene deforms neighboring hexagons and results buckling structure of stanene in the vicinity of the adsorption site, see Table I for detail.

\begin{figure}[t]
\includegraphics[width=0.5\textwidth,clip]{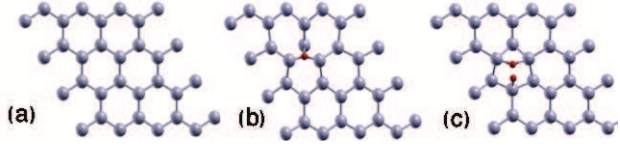}
\caption{Optimized structure of stanene (a) pristine, (b) oxygen atom adsorbed, and (c) oxygen molecule adsorbed.}
\end{figure}
\subsection{Tight binding calculations}
The tight binding model is based on an expansion of the electronic wave function into the basis of localized atomic wave functions. In the tight binding calculations we only consider the nearest neighbor atomic sites. The required tight binding parameters which include the on-site energy of tin atoms, hopping parameter between nearest neighbor sites and strength of spin-orbit interaction are extracting from our previous work \cite{R20}. For the pure 2D stanene we adopt single- and multi-orbital tight binding approximations. In the single- and multi-orbital models the intrinsic spin-orbit interaction is written as Kane-Mele term \cite{R14,R16,R20} and \textit{L.S} \cite{R20,R33}; respectively. In the Kane-Mele model, the intrinsic spin-orbit interaction term is written as an imaginary hopping between next-nearest neighbor sites. We applied the single orbital tight binding along with the Kane-Mele model for a zigzag edge stanene nano-ribbon with specific width. For a typical nano-ribbon, 1D periodicity is applied along the ribbon length and the ribbon width is limited to W atoms. The real space Hamiltonian matrix is Fourier transformed and diagonalized to find the electronic bands as a function of wave vector in the first Brillouin zone. The topological phase was examined under adsorption of oxygen atom and molecule on all different possible atomic sites in the unit-cell.  Here all tight binding calculations of 2D stanene and stanene nano-ribbon were performed by using a self-developed tight-binding code.

\section{Results and discussions}%

\begin{figure}[t]
\includegraphics[width=0.55\textwidth,clip]{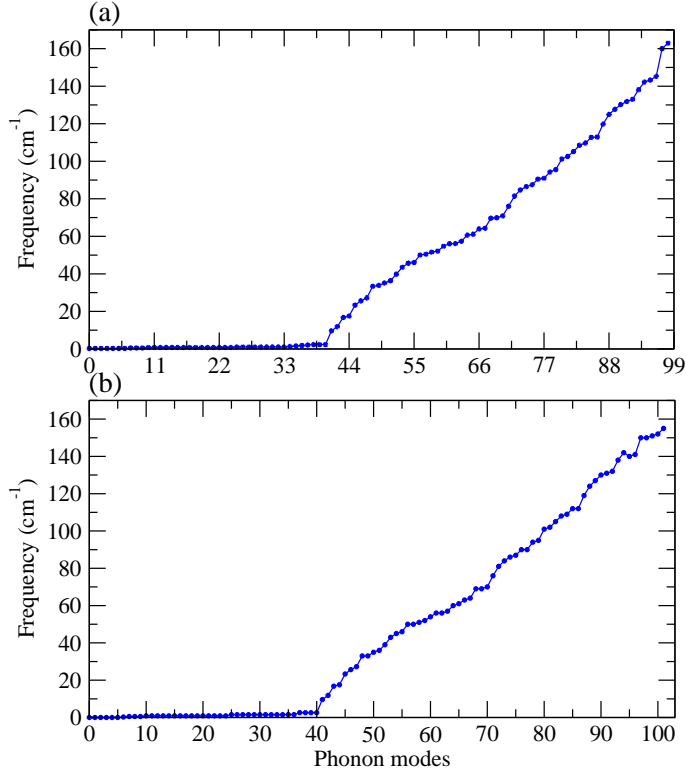}
\caption{$\Gamma$-point phonon frequencies for (a) oxygen atom adsorbed, and (b) oxygen molecule adsorbed stanene.}
\end{figure}
The stability of the adsorption of oxygen on the surface of 2D stanene is confirmed by performing $\Gamma$-point phonon calculations, see Fig.\ 2(a-b). From Fig.\ 2(a-b), it is confirmed that oxygen atom and oxygen molecule adsorbed stanene is stable because of the absence of negative phonon frequencies. This approach of evaluation of the structural stability has been applied for oxygen adsorbed graphene.\cite{carbon} Since each atom has three degrees of vibrational freedom and the O atom adsorbed stanene has 33 atoms in the system, therefore, the total vibrational modes in this system is 99, see x-axis in Fig.\ 2(a), while it becomes 102 for oxygen molecule adsorbed stanene because of 34 atoms in this system. The phonon frequency for pristine stanene has a value of $~180$ cm$^{-1}$ at the $\Gamma$-point \cite{scirep1}, however, this value is reduced to be 162 cm$^{-1}$ and 154 cm$^{-1}$ for oxygen atom and oxygen molecule adsorbed stanene (see Fig.\ 2(a-b)), respectively. This indicates that the vibrational frequencies are softened because of weakening of the Sn$-$Sn bonds due to the presence of the oxygen atom/molecule, which agrees well with previous reports for similar systems.\cite{carbon}

Note that the lattice parameter of the systems under consideration (18.68 \AA) is not modified by oxygen adsorption, while the bond lengths and buckling are affected, see Table I. Generally, in the vicinity of the adsorption site the Sn$-$Sn bond lengths and buckling are increased.
\begin{table*}[ht]
\begin{tabular}{|c|c|c|c|}
\hline
System & Sn$-$Sn & Buckling & Sn$-$O\\
 \hline
Stanene & 2.82 & 0.84 &--\\
\hline
Stanene$+$O & 2.83--3.70 & 0.65--2.69 & 2.20\\
\hline
Stanene$+$O$_2$ & 2.82--3.02 & 0.68--1.33 & 2.31\\
\hline
\end{tabular}
\caption{Bond lengths and buckling for the systems under consideration.}
\end{table*}

The main purpose of this study is to understand the effect of oxygen adsorption on the topological to trivial phase transition in zigzag stanene nano-ribbons. The oxygen adsorption determines one of the important environmental effects of future nano-devices based on topological insulators. DFT calculations for nano-ribbons are computationally expensive due to the larger unit cell of the one-dimensional structure. Hence, we adopt a method based on tight binding Hamiltonians from DFT calculations. We start with the DFT and tight binding calculations of two dimensional stanene. The required tight binding parameters for 2D stanene were reported in our previous work \cite{R20}. Fig.\ 3 shows the band structure of the $4\times4$ supercell of pristine stanene in both DFT and tight binding theories.
\begin{figure}[t]
\includegraphics[width=0.35\textwidth,clip]{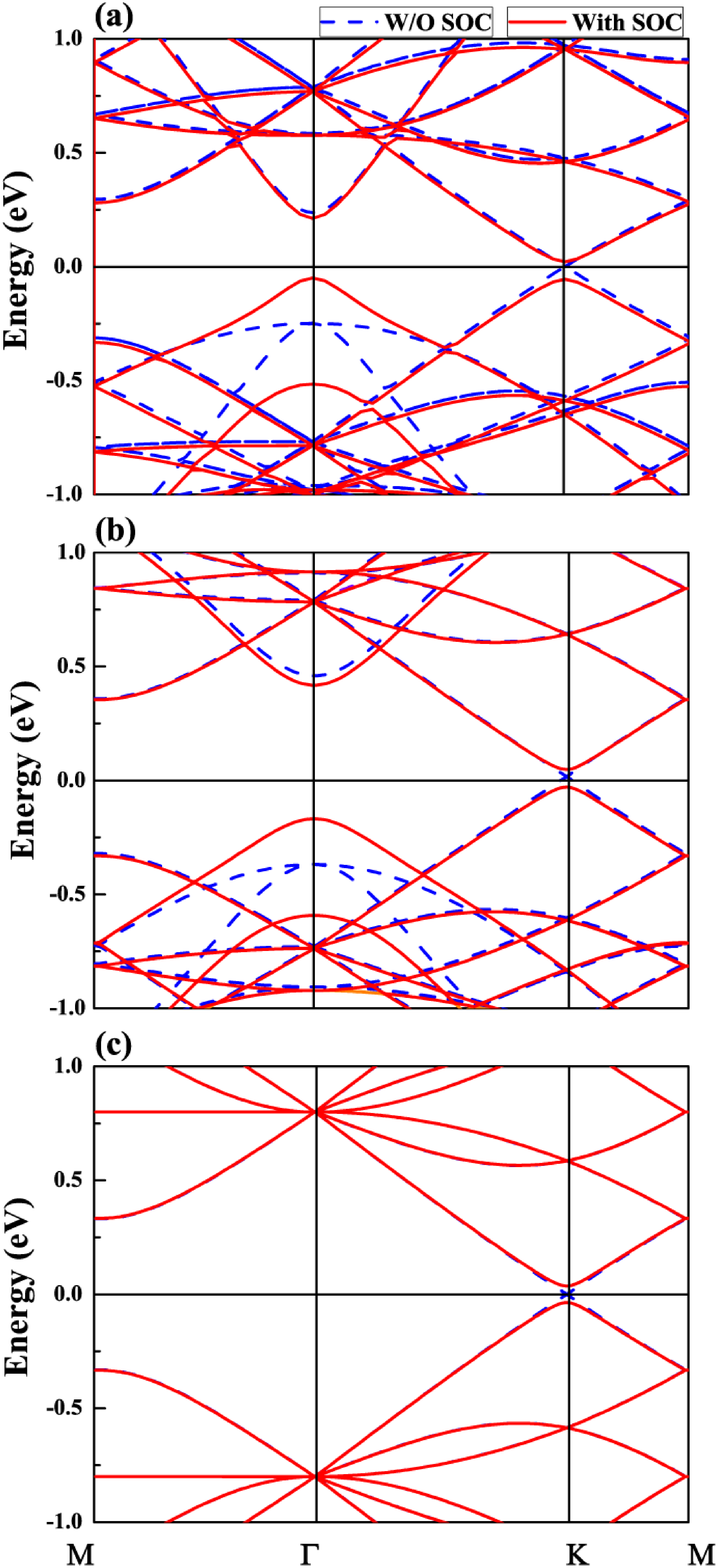}
\caption{Band structure of pristine stanene using (a) DFT calculations, (b-c) multi and single orbital tight binding models. The zero energy line represents the Fermi level.}
\end{figure}
In the absence of intrinsic spin-orbit interaction the electronic band gap of 2D stanene is zero at the K point. But both ab-initio and tight binding models predict an opening of the band gap at the K point of around 0.1 eV in pure stanene due to the strong spin-orbit interaction. The four atomic orbitals model describes the band structure in all region of the first Brillouin zone. The single orbital tight binding model does not match the DFT at the $\Gamma$ point but it describes the band structure around the Fermi level at the K point. The multi orbital model which includes the interaction between $\sigma$ and $\pi$ atomic orbitals solves the discrepancy at the $\Gamma$ point. To reduce the number of tight binding parameters and size of the Hamiltonian matrix for wide zigzag nano-ribbon calculations, we use the single orbital tight binding model in the remainder of paper. Additionally, one needs to understand the effect of oxygen adsorption on the tight binding parameters and Hamiltonian. We performed DFT calculations for adsorption of oxygen on the $4\times4$ supercell of 2D stanene. The DFT band structures for the optimized structures of oxygen atoms and molecules on stanene are plotted in Fig.\ 4(a).

\begin{figure}[t]
\includegraphics[width=0.35\textwidth,clip]{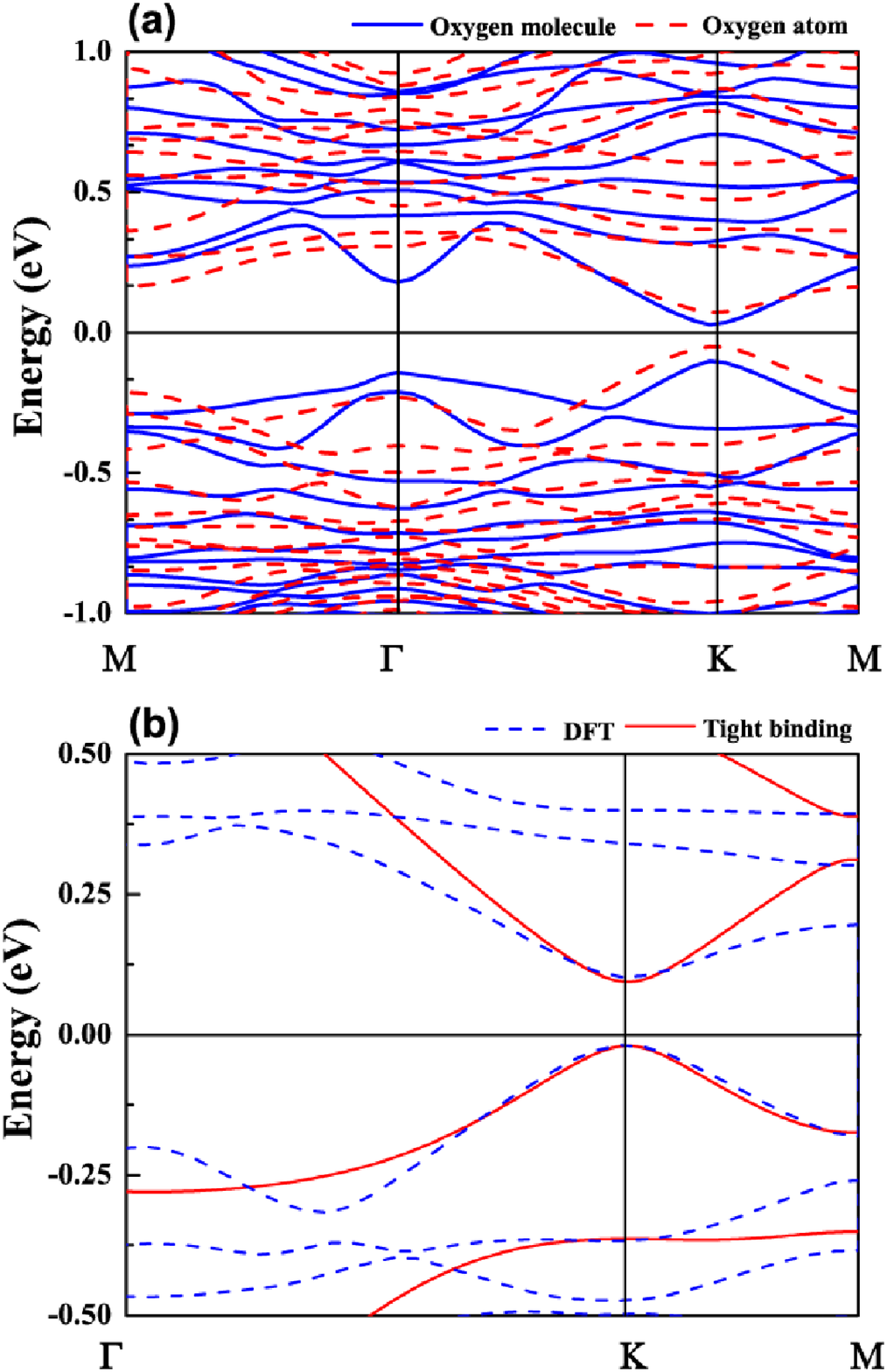}
\caption{(a) DFT band structure of oxygen atom and molecule adsorbed stanene and (b) low energy bands in the DFT and tight binding approximations for adsorption of a single oxygen atom on the 2D stanene.}
\end{figure}
In comparison to the pure stanene, adsorption of oxygen atom and molecule opens a direct electronic energy band gap of around 0.1 eV at the K point of the Brillouin zone. In both structures, the valence band maximum and conduction band minimum are located at the K point. The band gap in the functionalized stanene has potential application in nano-electronics. The band gap tuning by oxidation was reported for silicene experimentally \cite{R31} and theoretically \cite{t1,t2,t3}. It has been claimed that the free-standing silicene is unstable under oxygen adsorption due to the fact that oxygen molecules can  be dissociated into oxygen atoms in silicene with no overcoming energy barrier \cite{t3}, while, the energy barrier for an oxygen molecule indeed depends on the position and orientation of the overlying oxygen molecule \cite{t2}. Also the metallic state is gradually decaying due to oxygen adsorption in epitaxial silicene on Ag (111) \cite{R34}. For the tight binding study of oxygen adsorption on stanene, one needs the hopping parameters between tin and oxygen atoms. We match the DFT and single orbital tight binding results for adsorption of an oxygen atom on a 2D stanene as shown in Fig.\ 4(b). Fig.\ 4(b) compares the low energy band structure of stanene with a single oxygen atom in the DFT and single orbital tight binding models around the K point. In the fitting process we attend to the energy band gap and slope of low energy bands around the Fermi level. Although the two models do not match for high energy bands, the low energy states around the Fermi level at the K point are fairly well described in the tight binding approximation. By matching the two models, the on-site energy and hopping parameter between tin and oxygen atom are obtained as $-1$ eV and $-1.6$ eV, respectively. 

Using the above tight binding parameters we studied the adsorption of oxygen atoms on the zigzag edge stanene nano-ribbon. Fig.\ 5(a) shows the atomic structure and definition of unit-cell for zigzag stanene nano-ribbons. In the absence of O atoms, the structure is symmetric and the electronic properties of the ribbon are independent of the unit-cell length L. The smallest unit cell (L$=1$) has W atoms and is appropriate for pure zigzag nano-ribbons. Fig.\ 5(b) shows the electronic band structure of pure zigzag stanene nano-ribbon in the presence and absence of intrinsic spin-orbit interaction. In the absence of spin-orbit interaction, the low energy electronic states at the Fermi level are localized in the zigzag edges. The intrinsic spin-orbit interaction in the Kane-Mele model lifts the degeneracy of the edge states and produces helical edge states at the Fermi level. Between two TRIM points the helical edge state in the zigzag stanene nano-ribbon crosses the Fermi level in odd pairs which shows the topologically protected edge states as shown in Fig.\ 5(b). The band structure is plotted in first Brillouin zone between $\Gamma$ (k$=0$) and X (k$=$ $\pm$ $\pi$/a) points. The topological protected edge states are robust against weak disorder and interactions because time reversal symmetry prevents elastic back-scattering \cite{R28}.
\begin{figure}[t]
\includegraphics[width=0.55\textwidth,clip]{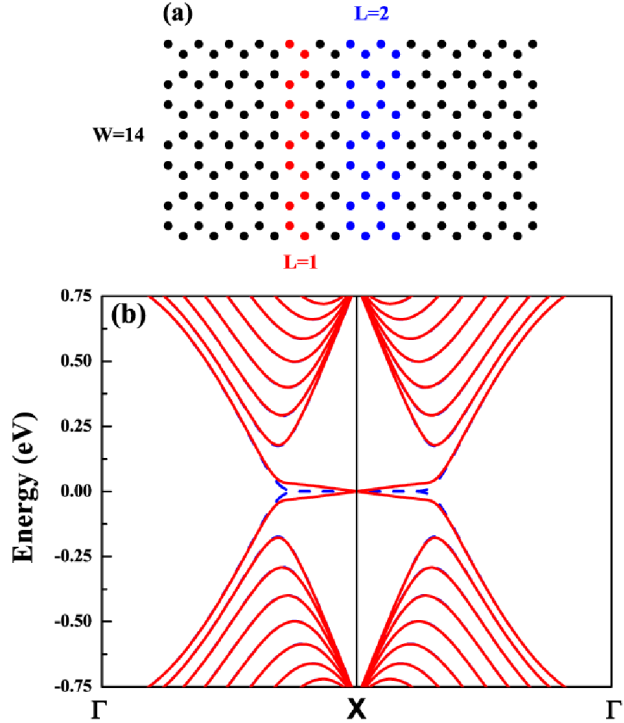}
\caption{(a) Atomic structure and unit-cell definition for zigzag stanene nano-ribbon (b) Low energy band structure of pure zigzag stanene nano-ribbon with W$=40$ atoms in the absence (blue dash) and presence (solid red) of SOC.}
\end{figure}
It was shown that the Hubbard term separates up and down spin states on the different zigzag edges of nano-structures and causes anti-ferromagnetic alignment of spins  on the edges \cite{R35,R36}. In the present work we ignore the electron-electron interaction and the electronic bands for spin up and down states are degenerate in the entire first Brillouin zone. It has been shown that the staggered sub-lattice potential, Rashba spin-orbit interaction \cite{R14} and random disorder \cite{R37} have the potential to remove the helical edge states and topological phase in zigzag stanene nano-ribbons. 

Our main goal is to study the effect of oxygen adsorption on the topological phase and helical edge states in the zigzag stanene nano-ribbons. We assume that the oxygen atoms are adsorbed on random atomic positions in the unit-cell of zigzag ribbon. The adsorption of the first O atom on the zigzag ribbon breaks the symmetry of structure. To model the exact adsorption process one should consider a unit-cell with large length which is computationally unavailable. Additionally L$=1$ leads to an oxygen chain along the ribbon length which amounts to an artificial order. To prevent the oxygen chain we also consider a unit cell with L$=$2 that has $2\times W$ atoms per unit-cell as shown in Fig.\ 5(a). We examined all different possible atomic sites for adsorption of a oxygen atom on the surface of the stanene nano-ribbon. Fig.\ 6(a-b) shows the band structure of the zigzag nano-ribbon after adsorption of a single oxygen atom on two different positions. 
 
\begin{figure}[t]
\includegraphics[width=0.7\textwidth,clip]{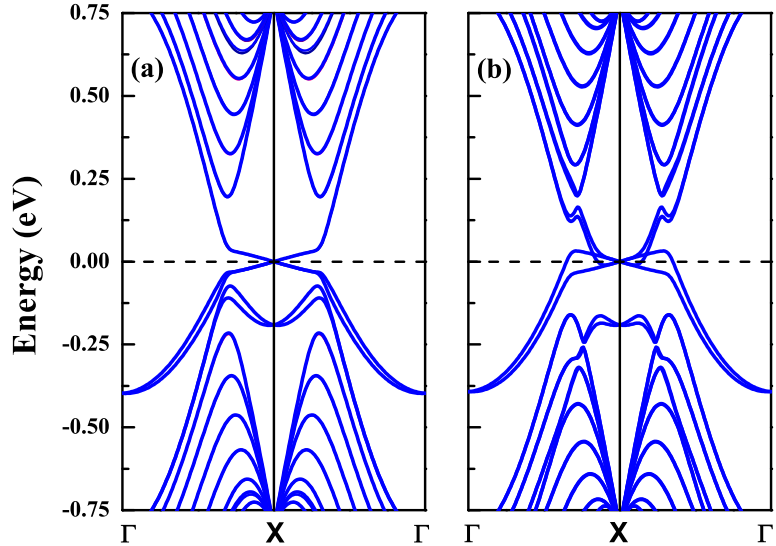}
\caption{(a-b) Electronic band structure of zigzag stanene nano-ribbon after oxygen adsorption on two different atomic sites for W$=40$ atoms and L$=1$.}
\end{figure} 
In Figs.\ 5(a) and 5(b), the electronic bands cross the Fermi level in odd/even pairs between TRIM points, which indicates the topological/trivial phase in the zigzag ribbon. After oxygen adsorption, both trivial and topological phases are possible as shown in Fig.\ 6(a-b). Based on our results, the atomic position of adsorbed oxygen atoms determines the topological/trivial phase of the zigzag stanene nano-ribbon after adsorption. We observe similar results for L$=$2 which are not presented here. For certain atomic positions, the adsorption of a single oxygen atom on the edge atoms conserves the topological phase of the zigzag stanene nano-ribbon. In all cases the oxygen atom conserves the helical edge state at the X point but additional crosses between energy bands and Fermi level occurs which determine the topological to trivial phase transition in zigzag stanene nano-ribbon. The adsorption of the next oxygen atoms on the surface of stanene nano-ribbons is a more complex problem. Generally for adsorption of N oxygen atoms on the zigzag ribbon with M atoms per unit-cell there are \ M!/N!(M-N)!\ different possible combinations. This number increases rapidly with ribbon width and number of oxygen atoms. The adsorption of more oxygen atoms will deform the energy bands more drastically and brings extra energy levels to the Fermi level. In the case of more oxygen atoms, the relative position of the oxygen atoms on the surface of the zigzag ribbon determines the number of crossings between energy bands and Fermi level and the trivial/topological phase of the ribbon. 
To specify the number of available topological structures, we define the topological phase probability as the number of topological structures divided by the number of all possible arrangement of oxygen atoms in the unit-cell. We examine all possible structures for L$=1$, 2 after adsorption of up to three oxygen atoms. The probability of the topological phase as a function of ribbon width is plotted in Fig.\ 7 for L$=1$, 2.

\begin{figure}[t]
\includegraphics[width=0.7\textwidth,clip]{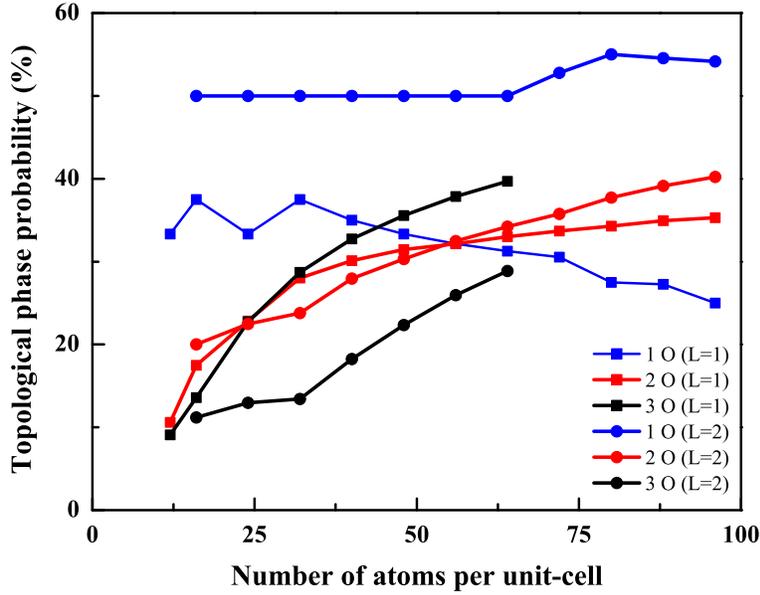}%(c) Topological probability of stanene with oxygen atom adsorption for different ribbon width and L$=$1 and L$=2$.
\caption{Topological probability of stanene with oxygen atom adsorption for different ribbon width for L$=$1 and L$=2$.}
\end{figure}
We consider up to three oxygen atoms on relatively wide zigzag nano-ribbons. For $L=2$ the number of tin atoms per unit-cell for a specific width is twice that of L$=$1. For higher numbers of oxygen atoms the calculations are very time-consuming and exceed our computational resources. Accordingly, they are left blank in Fig.\ 7. Generally, for L$=2$ the density of oxygen and the distortion of the band structure due to the adsorbed atoms is decreased which increases the possibility of the topological phase in the ribbon after oxygen adsorption. The possibility of the topological phase is almost width independent for adsorption of the first oxygen atom. For adsorption of two or three oxygen atoms the probability of the topological phase increases with ribbon width and tends to a constant value for wide enough zigzag nano-ribbons. The adsorption of additional oxygen atoms perturbs the band structure of narrow zigzag nano-ribbons and opens an energy band gap in the Fermi level which leads to the trivial phase. In the case of wider ribbons the effect of oxygen atoms is modified and the probability of the topological phase increases as shown in Fig.\ 7. The localized low energy edge states contribute effectively on the electronic structure of nano ribbon around the Fermi level. The adsorption of oxygen atom on the edge atoms perturbs the band structure in the Fermi level which may cause topological to trivial phase transition. On the other hand the topological phase is less sensitive to adsorption of oxygen on the non-edge atoms. According to Fig.\ 7, the topological phase and quantum spin Hall effect are sensitive to the adsorption of oxygen atoms but there is still a possibility for topological phase after oxidation of stanene zigzag edges. The topological phase in zigzag stanene nano-ribbons even after oxidation indicates that stanene has strong potential for experimental realization of spintronic nano-devices based on the topological phase of matter.

\section{Conclusion}
In summary, we present a study comprising DFT and tight binding approaches to investigate the adsorption of oxygen on the 2D structure and zigzag stanene nano-ribbons. By matching both models, we obtained the required tight binding parameters to study the topological phase in zigzag stanene nano-ribbons after oxygen atom adsorption. In addition, $\Gamma$-point phonon calculations were performed in order to make sure that the systems under study are stable. Our results demonstrate the possibility of the topological phase in zigzag stanene ribbons after oxidation which can be useful for future nano-electronics based on the topological phase of nano-materials.

\section{Acknowledgement}
G.S. and T.P.K. thank Prof. Michael Freund for his support. G.S. acknowledges funding from the Natural Sciences and Engineering Council of Canada (NSERC, Discovery Grant).

\end{document}